\documentclass[aip,amssymb,amsmath,reprint]{revtex4-1}
\usepackage[dvips]{graphicx}
\usepackage{tabularx}
\usepackage{color}
\usepackage[utf8]{inputenc}
\usepackage[T1]{fontenc}
\usepackage{mathptmx}
\graphicspath{{graf/}}

\begin{document}
\title{Transfer of a levitating nanoparticle between optical tweezers.}

\author{M. Calamai}
\affiliation{INFN, Sezione di Firenze, via Sansone 1, I-50019 Sesto Fiorentino (FI), Italy}

\author{A. Ranfagni}
\affiliation{INFN, Sezione di Firenze, via Sansone 1, I-50019 Sesto Fiorentino (FI), Italy}
\affiliation{European Laboratory for Non-Linear Spectroscopy (LENS), Via Carrara 1, I-50019 Sesto Fiorentino (FI), Italy}

\author{F. Marin}
\email[Electronic mail: ]{marin@fi.infn.it}
\affiliation{INFN, Sezione di Firenze, via Sansone 1, I-50019 Sesto Fiorentino (FI), Italy}
\affiliation{European Laboratory for Non-Linear Spectroscopy (LENS), Via Carrara 1, I-50019 Sesto Fiorentino (FI), Italy}
\affiliation{CNR-INO, L.go Enrico Fermi 6, I-50125 Firenze, Italy}
\affiliation{Dipartimento di Fisica e Astronomia, Universit\`a di Firenze, Via Sansone 1, I-50019 Sesto Fiorentino (FI), Italy}

\begin{abstract}
We demonstrate and characterize the transfer of a levitating silica nanosphere between two optical tweezers, at low pressure. Both optical traps are mounted on the heads of optical fibers and placed on translation stages in vacuum chambers. Our setup allows to physically separate the particle loading environment from the experimental chamber, where the second tweezer can position the particle inside a high Finesse optical cavity. The separation prevents from spoiling the cavity mirrors and the chamber cleanliness during the particle loading phase. Our system provides a very reliable and simply reproducible protocol for preparing cavity optomechanics experiments with levitating nanoparticles, opening the way to systematic studies of quantum phenomena and easing the realization of sensing devices.       
\end{abstract}

\maketitle

\section{Introduction}

Quantum optomechanics has recently expanded the range of explored and exploited systems to nanoparticles levitating in vacuum, trapped and oscillating in the potential created by an optical field\cite{Chang:2010,Barker:2010,Romero:2010,Millen:2020}. In particular, the topic of cavity optomechanics is very intriguing for the possibility of realizing quantum coupling between photonic field and the particle motion\cite{Romero:2011}, where the latter is strongly decoupled from environmental thermal noise by operating in high vacuum. 

Most proposals\cite{Chang:2010,Barker:2010,Romero:2010} and experiments\cite{Kiesel:2013,Asenbaum:2013,Millen:2015,Meyer:2019} aiming to cool the dynamics of a levitating nanoparticle inside an optical cavity are based on the dispersive coupling 
of its motion to the electromagnetic field, a technique well investigated in optomechanics\cite{Aspelmeyer:2014_review}.
A different mechanism of cavity cooling, relying on coherent trapping of light scattered by a
levitating nanoparticle  into an optical cavity, has been recently 
realized \cite{Widney:2019,Delic:2019_A} and allowed to achieve motional
cooling of a levitated nanoparticle to a  phononic occupation number below unity \cite{Delic:2020}. In any case, accurate positioning of the nanoparticle inside the cavity is crucial to tune and optimize the optomechanical coupling.  

To optically trap a neutral nanoparticle, a laser beam is tightly focused 
in a chamber, in the presence of gas containing suspended particles.
If their motion is sufficiently damped by collisions with the background gas,
trapping occurs as one particle crosses the focused beam and releases its kinetic energy fast enough to be captured by the optical potential.
To implement cavity optomechanics experiments, it is then necessary to place the levitating particle into the region defined by a field mode of a high finesse optical cavity, with sub-micrometric precision. The position must be stably and accurately maintained, avoiding excess mechanical and acoustic vibrations. 
A prerequisite is loading the dipole trap (optical tweezer) without spoiling the cavity mirrors, something that easily occurs due to particle deposition on the mirrors surface. 
Finally, high vacuum conditions must be achieved in reasonable time, maintaining stable conditions. Even this latter procedure is conditioned by the relatively high pressure necessary for the initial trapping stage, and often by the presence of solvents used for injecting the particles in the chamber through a nebulizer.     
A clean and reproducible method to prepare a levitating nanoparticle for cavity optomechanical experiments is actually very useful, but not straightforward. 

A possibility is loading the particle on the optical tweezer in a first chamber, than transfer it to a cleaner environment containing the optical cavity and the positioner. 
A movable optical trap is described in Ref. \onlinecite{Mestres:2015}. The trap is loaded in a first chamber using a nebulizer, then the whole tweezer, mounted on micrometric positioners in an extensible arm, is moved to a second chamber and the particle is delivered to the stationary wave of an optical cavity. 
To stabilize the particle during the transfer, a cooling scheme acting on the tweezer optical power is used.
A different method to transfer a levitating particle between different vacuum chambers is described in Ref. \onlinecite{Grass:2016}. A standing wave is created inside a hollow fiber connecting the two chambers, by means of counter-propagating laser beams. The particle is trapped on an anti-node of the standing wave, then moved by slightly shifting the two beams frequencies. The collection of the particle in the second chamber has not yet been reported. 

In this work we describe a method for reliably  
loading a nanoparticle on a stable and accuratly positionable tweezer, inside a high finesse optical cavity, avoiding mirrors performance degradation. Similarly to the work of Ref. \onlinecite{Mestres:2015}, the particle is trapped in a first chamber by a tweezer placed on a movable arm, then translated into the experimental chamber containing the optical cavity. It is then transferred to a second optical trap, that is mounted inside the second chamber on nano-positioners. 
This second tweezer is used to accurately position the particle inside the optical cavity. 
Mounting the nano-positioners on the chamber basement that also support the optical cavity, instead of placing them on the moving arm, significantly improves the overall mechanical stability. Moreover, the moving arm is retracted after the particle transfer, and the vacuum chambers isolated. As a consequence, the environment in the experiment chamber is suitable for a rapid evacuation down to very low pressure.

The crucial stage in our scheme is the transfer of the nanoparticle between two optical tweezers, in a low pressure environment. In the following we characterize in detail this procedure. 

\section{Experiment}

\begin{figure}
\includegraphics{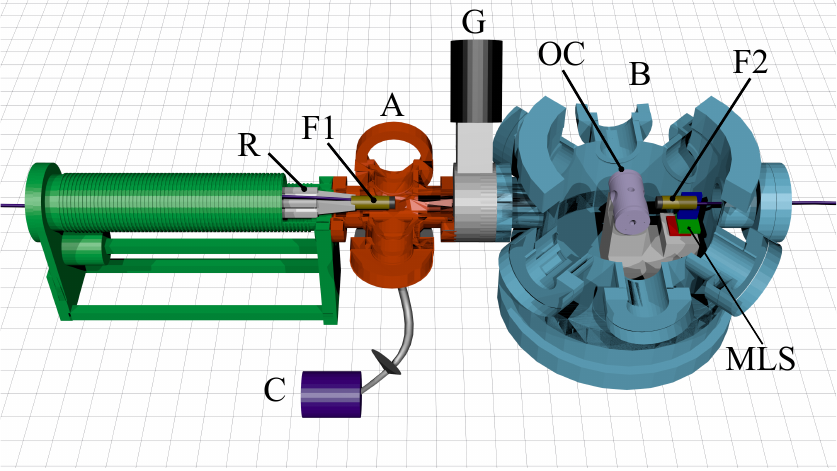}
\caption{Experimental setup. Nanoparticles are injected in chamber
C, then transported in a gas flux toward chamber A where they are captured by the tightly focused light of a laser diode delivered by a single-mode fiber. The fiber head with the focusing system (F1)  
is mounted on the tip of a rod (R) that can be manually translated 
between chambers A and B through the gate G.
The second optical tweezer is formed by the light of a Nd:YAG laser, delivered by a similar optical system
(F2) mounted on a three-axis miniature linear translation
stage (MLS). The focus can be positioned inside the optical cavity (OC) with sub-micrometric precision.}
\label{setup}
\end{figure}

The setup is shown in Fig. \ref{setup}. 
Nanoparticles are caught in chamber A, 
then transferred to the second trap in chamber B.
The optical tweezer that capture the particles is realized with a fibered 976 nm 
laser diode (LD). The light delivered by a single-mode fiber 
is collimated 
and focused using an optical system (F1) composed of 
two aspheric lenses, having nominal focal length and numerical aperture of respectively 15.4 mm (N.A. 0.16) and 3.1 mm (N.A.  0.68). 
The two lenses are screwed on the fiber head connector. The beam at the focus is elliptical with waists of $0.96 \,\mu$m and $0.92 \,\mu$m, as deduced from the particle oscillation frequencies at the typical output power of 250 mW.
The fiber head with the optics is mounted at the end of a 500 mm long, X-shape aluminum rod screwed on the moving flange of a bellowed sealed linear shift mechanism (HV Design) that allows  
to manually translate it between 
chambers A an B.
We note that this support is sensitive
to mechanical vibrations, making this trap
unsuitable for stable cavity optomechanics experiments. 

A drop of aqueous 
solution of silica nanospheres (9\% of particles, in mass) of radius $\sim$85 nm 
is injected 
inside chamber C, that is filled 
with clean nitrogen while chamber A is evacuated. The valve separating the 
two chambers is opened and dust of nanoparticles is
introduced in chamber A, carried by the gas turbulence produced by 
the pressure unbalance.
Trapping by the optical tweezer occurs when a pressure of  $\sim$100 mbar is achieved in chamber A, typically within few minutes.

With a particle trapped, before opening 
the gate G, residual wandering nanoparticles are 
pumped out from chamber A, whose pressure is gently decreased down to the  
mbar level. The chamber is then  
slowly refilled with pure nitrogen up to $\sim$30 mbar, and the gate is opened to equilibrate the pressure between chamber A and
B. The optical tweezer is translated to chamber B and positioned in front of the second optical trap.
We remark that at this pressure the
nanoparticle motion is over damped, and we can keep the levitating particle during the translation without using any active feedback.

The second tweezer is formed by the 1064 nm radiation of a Nd:YAG laser, delivered into chamber B by a polarization  maintaining fiber. The focusing optical system screwed on the fiber head (F2) is the same of the first tweezer, and is positioned on a
three-dimensional miniature linear stage
(PI Q-522). The beam waists at the focus are $1.02 \mu$m and $0.93 \mu$m, the typical optical power is 200 mW. Fibered beam-splitters allow to collect part the light arriving from the fiber heads. With the help of dichroic mirrors, we can thus measure the transmitted and back-scattered light of both sources. 

\begin{figure}
\includegraphics{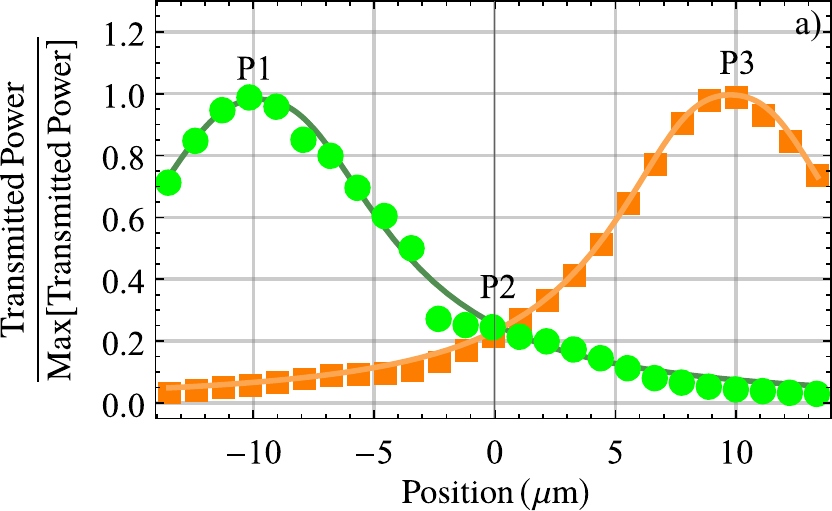}
\includegraphics{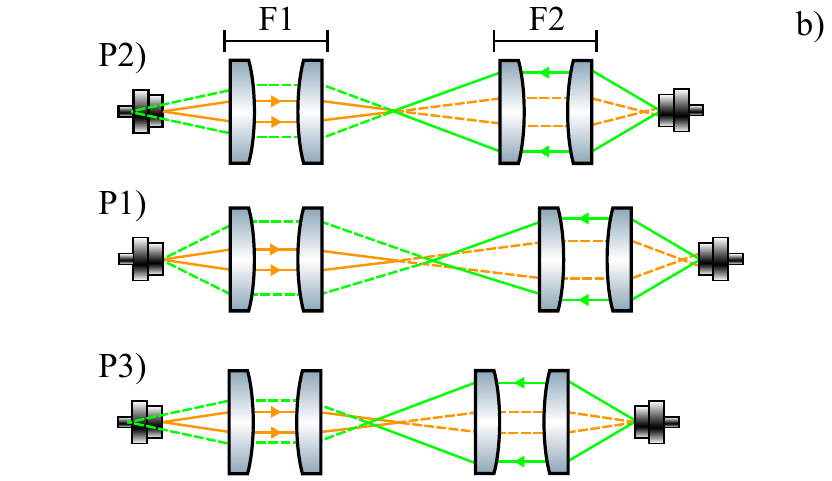}
\caption{(a) Transmitted power of the laser light from the two sources through the two optical fibers and the corresponding F1 and F2 optical systems. Green dots: Nd:YAG light. Orange squares: LD light. Data are recorded approaching 
the two fiber heads at $1.1 \,\mu$m per step, and normalized to the maximum transmitted power 
for each wavelength. Abscissa represents the variation of fiber heads distance,
with the origin set halfway between the two maxima. Solid lines: overlap integral between the propagating field modes, fitted to the experimental data. (b) Schematic drawing 
of the two focusing systems during the measurement. 
Green (orange) rays represent the Nd:YAG (LD) beam 
propagation, with arrows indicating the direction. 
P2 indicate the optimal position to transfer the particle, as the two focuses
are spatially overlapped.
At relative position P1(P3) the two focusing systems
are optimally placed to couple the Nd:YAG (LD) optical power.
In that case, the distance between the two traps is 9.8 $\mu$m.
 }
\label{aberrazione}
\end{figure}

To transfer the particle between the two tweezers, we have to superpose the positions of their intensity maxima with sub-micrometric precision. This procedure is performed by moving the second fiber head. Its transverse position with respect to the optical axis is optimized by maximizing the light transmission between the two fibers, while the distance between the two fiber heads must take into account the chromatic aberration, as sketched in Fig. \ref{aberrazione}(b).
We remark that the light of the second tweezer remains off during the whole procedure, to avoid the accidental formation of an unstable potential by the superposition of the two intensity profiles. 

To define the optimization procedure, we have performed a preliminary characterization of the optical coupling between the two fibers, at the two used wavelenghts. 
The transmitted power of the Nd:YAG light through the first fiber, and that of the LD light through the second fiber are reported in Fig. \ref{aberrazione}(a). The transverse position of the fiber head is kept optimized during the measurement, while the two fiber heads are moved closer  
at $\sim 1.1 \mu$m steps. 
The solid line, for each of the two wavelengths, is 
given by the overlap integral of the two counter-propagating modes, fitted to the experimental data.

We find a distance of
9.8 $\mu$m between the positions of the foci for the two wavelengths.
As shown in the scheme of Fig. \ref{aberrazione}(b), assuming two identical focusing 
systems the optimal distance to transfer the particle between the tweezers is
halfway between the transmission maxima at the two wavelengths (this position is labeled as P2 in the figure).
The operative procedure is then the following: we optimize the transmission of the LD light through the second fiber by moving the fiber head in the three directions, and afterwards we increase the fiber heads distance by $\sim 10 \mu$m.

\begin{figure}[htbp]
{\includegraphics[width=0.99\columnwidth]{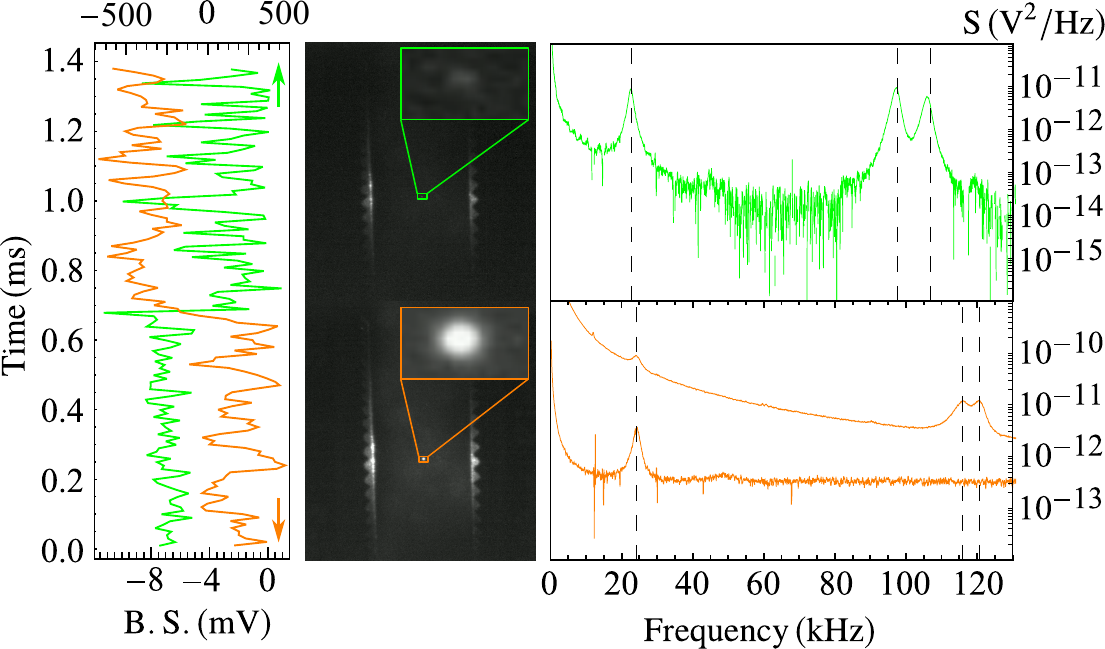}}
\caption{Left panel: the light back scattered (B.S.) from the the particle is collected from the trapping fibers
during the transfer between the LD and the Nd:YAG tweezers. Orange: LD light signal (scale on the bottom axis). Green: Nd:YAG light signal (scale on the top axis).
Central panel: images of the Nanoparticle trapped by the LD (bottom picture) and the Nd:YAG (top picture) optical tweezers.
Bright spots, also shown in the enlarged insets, are due to the particle dipole emission, and scattered light allows to identify the edges of the focusing lenses. Brightness difference between the two traps 
is due to the different camera sensitivity
at the two wavelengths. Right panel: spectra of the back and forward scattered light, collected by the fibers and acquired at a background pressure of 2 mbar, exhibiting spectral peaks corresponding to the three eigenfrequencies of the particle motion. 
Bottom graph: spectra of the forward scattering (upper trace) and back scattering (lower trace) of the LD light, with the particle on the first tweezer. Top graph: spectrum of the forward scattering of the Nd:YAG light, with the particle trapped by the second tweezer. Vertical dashed lines display the particle oscillation frequencies.
}
\label{two_traps}
\end{figure}
To load the second trap, we boost the Nd:YAG power 
and slowly turn off the LD. 
With the described protocol, we can reliably transfer the 
particle between the two traps. In Fig. \ref{two_traps} we show 
a photo of the two optical systems and the levitating nanoparticle 
before and after the transfer.
The power spectra of the light collected by the fibers in the back and forward directions, also shown in Fig. \ref{two_traps}, exhibit the peaks associated to the  nanoparticle motion in the three orthogonal directions defined by the trap geometry. In both cases, the background pressure is reduced down to 2 mbar to show such clear signatures of the under-damped motion.  

We often observe that the particle scattered light 
changes suddenly during the transfer. On the other hand, a transfer between potential wells having the same minimal point should be characterized by a continuous change in the apparent particle brightness, following the varying light intensity. The observed abrupt changes indicate that the nanoparticle jumps between two potential minima, that are not perfectly superimposed due to an uncertainty in the positioning of the order of few hundred nanometers, and to the optics mechanical vibrations. 
While turning off the LD, the potential barrier from the first to the second trap, as well as the depth of the first trap, become vanishing small. A reliable transfer occurs if the jump rate (favored by the lowering barrier) is higher enough than the loss rate (increased by the lowering well depth) and the motion is damped enough that the particle can loose its kinetic energy during the transfer. 
Our double tweezer is a versatile system to investigate the stochastic motion of a particle in a variable three-dimensional potential\cite{Rondin:2017}, that is however beyond the scope of the present article and is left to further works.  
At the purpose of providing useful information for the reproduction of our method, we describe in the following a semi-quantitative investigation of the pressure and misalignment ranges that allow a reliable transfer.

\begin{figure}
\includegraphics{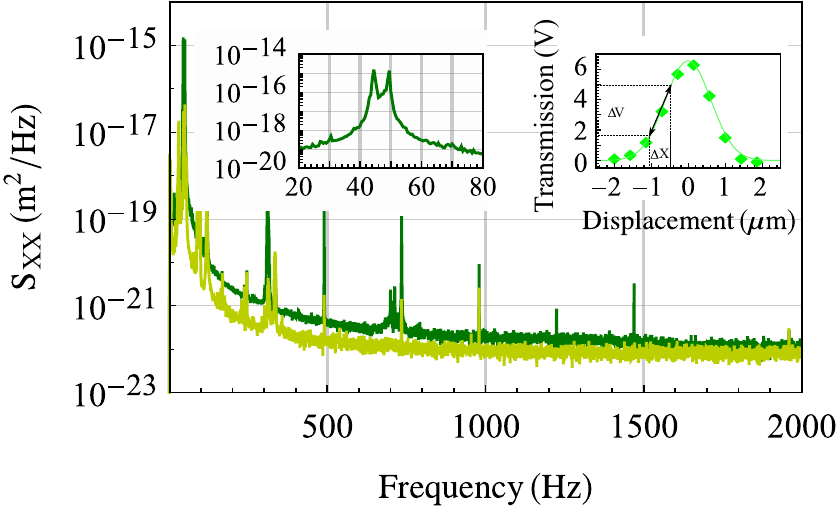} 
\caption{Spectra of the relative position between the two fiber heads, on the plane perpendicular to the tweezer axis, along the vertical (green) and horizontal (yellow)
directions. Left inset: dominant vibrational modes at 44 Hz and 49 Hz in the vertical direction.
Right inset: transmitted power of the Nd:YAG light through the two optical fibers, as the second fiber head is translated in the vertical direction. This curve is used to convert into displacement spectra the acquired transmission spectra, as illustrated in the picture.
}
\label{vibrazione}
\end{figure}

We first characterize the relative mechanical vibrations of the two trapping optics  
on the plane perpendicular to the optical axis.
The two focusing systems are first placed at the position that maximizes
the transmitted Nd:YAG power through the two fibers. The transmitted signal is then recorded 
while moving the second fiber head in the vertical direction. Hence, the fiber head is set at the position that halves the transmitted power, and the time trace of the transmitted signal is acquired and calibrated in terms of displacement fluctuations using the previously recorded transmission curve (as illustrated in the right inset of Fig. \ref{vibrazione}). The same procedure is repeated for the horizontal displacement. In Fig. \ref{vibrazione} we show the calculated displacement noise spectra.
The main spectral feature is a double peak at $\sim$ 50 Hz for the vertical
direction, whose area corresponds to a displacement of $\sim 50$ nm (root means square), much smaller than the beam waists.  A simulation with a Finite Element Model 
shows that the two peaks are due to flexural modes of the rod that sustains the first fiber head.

In order to define the pressure range that allows a reliable transfer, we have repeated at least three times the transfer back and forth between the two traps, 
at the pressure values of 100, 75, 50, 25 and 15 mbar. We actually lost the particle during the fourth attempt at 10 mbar. We notice that at 10 mbar the damping rate is about $\Gamma \simeq 2 \pi \times 10 \,$kHz, thus the particle motion is weakly damped.

At 50 mbar we have then evaluated the tolerance in the misalignment between the two fiber heads. Starting from the optimal position, 
we could transfer the particle three times back and forth in different relative positions, until the two focuses 
were misplaced by $\sim 3 \, \mu$m on the plane
perpendicular to the optical axes, or  $\sim 10 \, \mu$m along the optical axes. For the latter case, we show in Fig. \ref{two_traps}, on the left panel, the time evolution of the backscattered light during a transfer from the LD to the Nd:YAG tweezers. The visible steps indicate a jump between the two potential wells occurring in a time shorter than $0.1 \, $ms. 

\begin{figure}
{\includegraphics{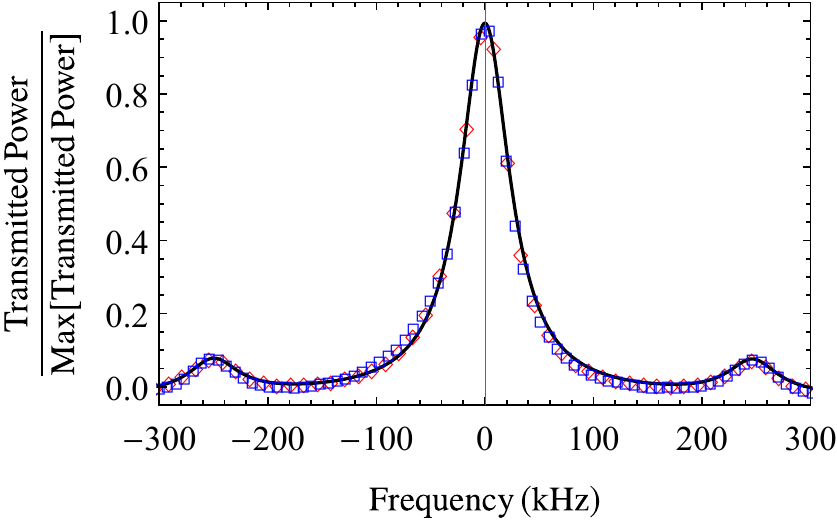}} 
\caption{Transmitted power of a probe Nd:YAG laser through the high Finesse optical cavity. Sidebands at $\pm 250 \,$kHz are produced by laser phase modulation for calibrating the frequency scan. The blue squares and red diamonds correspond to
acquisitions recorded respectively before the nanoparticle capture and transfer, and after ten complete operations. The solid line shows a fit to the latter data set.
}
\label{finesse}
\end{figure}

After having defined the above described transfer protocol, we have placed a $\sim 50 \,$mm long optical cavity 
(Finesse 54000) inside the chamber B. The cavity spacer has a 20 mm diameter radial hole that allows to place on the cavity optical axis the nanoparticle trapped by the Nd:YAG tweezer. 
We have captured and transferred several particles 
from the LD to the Nd:YAG trap at a background pressure of 30 mbar.
Even after ten complete cycles we could appreciate no 
degradation of the cavity finesse, as shown in Fig. (\ref{finesse})
where we report two recordings of the cavity transmission function, acquired before the first, and
after the last transfer operation.
The measured width is respectively $57 \pm 1$ kHz and $56 \pm 1.5$ kHz.

\section{Conclusions}

We have described a robust method to systematically capture and place a levitated
nanoparticle with sub-micrometric precision inside an optical cavity,
without spoiling the cavity optical quality and the environment of the experimental chamber.  
We have indeed observed that the optical cavity can be repeatedly loaded without degrading its performance. A key element of our protocol is the transfer of the loaded particle between two, completely independent optical tweezers, that we realize and characterize, and that is performed without the necessity of stabilizing feedback loops. Hence, the second tweezer can be mounted on stable nano-positioners, that allow a reliable and systematic investigation of the coupling between the nano-particle and the cavity optical field. Our system provides a very reliable and simply reproducible protocol for preparing cavity optomechanics experiments with levitating nanoparticles, opening the way to systematic studies of quantum phenomena and easing the realization of sensing devices\cite{Ranjit:2016}. 

\begin{acknowledgments}
We thank A. Pontin for the assistance at the early stage of this work. Research performed within the Project QuaSeRT funded by the QuantERA ERA-NET
Cofund in Quantum Technologies implemented within the European Union's
Horizon 2020 Programme. 
\end{acknowledgments}

\section*{DATA AVAILABILITY}
The data that support the findings of this study are available from the corresponding author
upon reasonable request.

\bibliographystyle{apsrev4-2} 
\bibliography{references}

\end{document}